\documentclass[aps, superscriptaddress, twocolumn, prb]{revtex4-2}
\usepackage{amsfonts}
\usepackage{amssymb}
\usepackage{amsmath}
\usepackage{graphics}
\usepackage{graphicx}
\usepackage{subfigure}
\usepackage{dcolumn}

\usepackage{bm}
\usepackage{color}
\usepackage{epstopdf}

\definecolor{Gray}{gray}{0.1}

\DeclareMathAlphabet\mathbfcal{OMS}{cmsy}{b}{n}

\begin{document}

\title{Intense anomalous high harmonics in graphene quantum dots caused by
disorder or vacancies}
\author{H.K. Avetissian}
\author{G.A. Musayelyan}
\author{G.F. Mkrtchian}
\thanks{mkrtchian@ysu.am}

\affiliation{Centre of Strong Fields Physics at Research Institute of Physics, Yerevan State University,
Yerevan 0025, Armenia}
\begin{abstract}
This article aims to study the linear and nonlinear optical response of
inversion symmetric graphene quantum dots (GQDs) in the presence of on-site
disorder or vacancies. The presence of disorder or vacancy breaks the
special inversion symmetry leading to the emergence of intense Hall-type
anomalous harmonics. This phenomenon is attributed to the intrinsic
time-reversal symmetry-breaking in quantum dots with a pseudo-relativistic
Hamiltonian, even in the absence of an external magnetic field. We
demonstrate that the effects induced by disorder or vacancy have a distinct
impact on the optical response of GQDs. In the linear response, we observe
significant Hall conductivity. In the presence of an intense laser field, we
observe the radiation of strong anomalous odd and even-order harmonics
already for relatively small levels of disorder or mono-vacancy. The both
disorder and vacancy lift the degeneracy of states, thereby creating new
channels for interband transitions and enhancing the emission of near-cutoff
high-harmonic signals.
\end{abstract}

\maketitle

\section{Introduction}

Theoretical studies of two-dimensional (2D) systems with unique electronic
and topological properties can be traced back about 80 years, encompassing
areas such as single graphite layers \cite%
{wallace1947band,semenoff1984condensed}, zero-gap semiconductors \cite%
{fradkin1986critical}, the quantum Hall effect without a magnetic field \cite%
{haldane1988model}, d-wave superconductors \cite{lee1993localized}, and
neutrino billiards \cite{berry1987neutrino}. However, it was only with the
experimental realization of graphene in 2004 \cite{novoselov2004electric}
that a new frontier in physics opened up: the field of 2D materials
featuring pseudo-relativistic charged carriers and nontrivial spatial and
band structure topology \cite%
{neto2009electronic,hasan2010colloquium,sarma2011electronic,qi2011topological,manzeli20172d}%
. One remarkable manifestation of the importance of topology is the "mystery
of a missing pie" \cite{geim2007rise} in the DC conductivity of graphene $%
\sigma _{DC}$. Early experiments observed values of $\sigma _{DC}$ that were 
$\pi $ times larger than the predicted value for pristine, disorder-free
graphene \cite{shon1998quantum}. Subsequent studies revealed that the value
of $\sigma _{DC}$ strongly depends on various factors, including the
boundary conditions of the graphene sheet \cite%
{tworzydlo2006sub,miao2007phase}, the presence of disorder\cite%
{ando2002dynamical,ostrovsky2006electron}, and the interactions among
charged carriers \cite%
{mishchenko2007effect,herbut2008coulomb,rosenstein2013chiral}. This
highlighted the intricate interplay between topology, disorder, and carrier
interactions in graphene's electrical conductivity.

The optoelectronic properties of graphene undergo significant changes when
it is reduced to zero-dimensional structures known as graphene quantum dots
(GQDs) \cite{gucclu2014graphene}. The behavior of GQDs can vary from
metallic to insulating, depending on the type of edges they possess \cite%
{zarenia2011energy}. The optical properties of GQDs depend on their size and
shape \cite%
{yamamoto2006edge,ozfidan2014microscopic,gucclu2014graphene,fang2014active,pohle2018symmetry}%
.

Various synthesis methods exist for GQDs, including fragmentation of
fullerene molecules \cite{lu2011transforming} and nanoscale cutting of
graphite combined with exfoliation \cite{mohanty2012nanotomy}, decomposition
of hydrocarbons \cite{olle2012yield}. The confinement of electronic states
in GQDs has been confirmed through scanning tunneling microscopy
measurements \cite{ritter2009influence}. Along with graphene, GQDs have
garnered significant attention due to their unique optoelectronic properties
and their potential applications in diverse fields such as bioimaging \cite%
{zhu2011strongly,younis2020recent}, photovoltaics \cite%
{li2011electrochemical}, quantum computing \cite{trauzettel2007spin},
photodetectors \cite{kim2014high}, energy storage \cite{zhai2022carbon},
sensing \cite{li2019review}, and metal ion detection \cite{ting2015graphene}.

The potential to engineer the energy spectrum and optical transitions of
GQDs has got significant attention also in the field of strong-field physics 
\cite{avetissian2015relativistic}. This is due to the promising prospects
for GQDs in extreme nonlinear optics applications, including high-order
harmonic generation (HHG) \cite{corkum1993plasma}. Theoretical studies have
predicted strong HHG from fullerenes \cite%
{zhang2005optical,zhang2006ellipticity,zhang2020high,avetissian2021high,avetissian2023disorder}%
, graphene nanoribbons \cite%
{cox2017plasmon,avetissian2020extreme,zhang2022extended}, and GQDs \cite%
{JETP2022high,JN2022high,JETPL2022laser,gnawali2022ultrafast,
avetissian2023long}. These studies have also highlighted significant
alterations in the nonlinear optical properties by manipulating the size,
shape, and edges of these systems. It is worth noting that the
electron-electron interactions have been recognized as a crucial factor that
influence on the optical phenomena in the mentioned nanostructures \cite%
{avetissian2023long}.

In most previous theoretical studies, the interaction of graphene-like
systems with the laser fields has been primarily focused on perfect crystal
structures with periodic lattices. However, as is known disorder or point
defects in graphene, such as vacancies can strongly affect the electronic
properties of the system, since those defects support quasi-localized
electronic states \cite{pereira2006disorder,palacios2008vacancy,nair2012spin}%
. Recent studies have demonstrated that an imperfect lattice can lead to
enhancement of HHG compared to a perfect lattice, particularly when
considering doping-type impurities or disorder \cite%
{yu2019enhanced,yu2019high,pattanayak2020influence,hansen2022doping,avetissian2023disorder}%
. This raises the question of how the disorder or vacancies specifically
affect the HHG spectra in GQDs. It is worth noting that GQDs with sharp
boundaries exhibit time-reversal symmetry-breaking even in the absence of a
magnetic field \cite{berry1987neutrino}, suggesting the possibility of
Hall-type anomalous responses in both linear and nonlinear regimes for
imperfect GQDs. In this study, we present a microscopic theory that explores
the linear and nonlinear interaction of GQDs with strong coherent
electromagnetic radiation in the presence of on-site disorder or vacancy.
Specifically, we investigate hexagonal GQD with a moderate size, as
illustrated in Figure 1. By employing up to 10th nearest-neighbor hopping
integrals within a dynamical Hartree-Fock (HF) approximation, we reveal the
polarization-resolved structure of the HHG spectrum.

The paper is organized as follows. In Sec. II, the model with the basic
equations are formulated. In Sec. III the linear optical response is
considered. In Sec. IV, we present the results regarding nonlinear optical
response. Finally, conclusions are given in Sec. V.

\section{The model}

The basic GQD$_{216}$, illustrated in Fig. 1, has symmetry described by the
non-abelian point group C$_{6v}$. We also consider GQD$_{216}$ with on-site
disorder and GQD$_{216}$ with a mono-vacancy. GQD is assumed to interact
with a laser pulse that excites electron coherent dynamics. We assume a
neutral GQD that will be described in the scope of the TB theory. Hence, the
total Hamiltonian reads:
\begin{equation}
\widehat{H}=\widehat{H}_{\mathrm{TB}}+\widehat{H}_{\mathrm{C}}+\widehat{H}_{%
\mathrm{int}},  \label{Ht}
\end{equation}%
where%
\begin{equation}
\widehat{H}_{\mathrm{TB}}=\sum_{i\sigma }\varepsilon _{i}c_{i\sigma
}^{\dagger }c_{j\sigma }-\sum_{i,j\sigma }t_{ij}c_{i\sigma }^{\dagger
}c_{j\sigma }  \label{HTB}
\end{equation}%
is the free GQD TB\ Hamiltonian. Here $c_{i\sigma }^{\dagger }$\ ($%
c_{i\sigma }$) creates (annihilates) an electron with the spin polarization $%
\sigma =\left\{ \uparrow ,\downarrow \right\} $\ at the site $i$\ ($%
\overline{\sigma }$\ is the opposite to $\sigma $\ spin polarization). In
Eq. (\ref{HTB}) $\varepsilon _{i}$ is the energy level at the site $i$, and $%
t_{ij}$ is the hopping integral between the sites $i$ and $j$.

The second term in the total Hamiltonian (\ref{Ht}) describes the
electron-electron interaction (EEI):%
\begin{equation}
\widehat{H}_{\mathrm{C}}=\frac{U}{2}\sum_{i\sigma }c_{i\sigma }^{\dagger
}c_{i\sigma }c_{i\overline{\sigma }}^{\dagger }c_{i\overline{\sigma }}+\frac{%
1}{2}\sum_{i,j\sigma \sigma ^{\prime }}V_{ij}c_{i\sigma }^{\dagger
}c_{i\sigma }c_{j\sigma ^{\prime }}^{\dagger }c_{j\sigma ^{\prime }},
\label{c}
\end{equation}%
with the parameters $U$ and $V_{ij}$ representing the on-site, and the
long-range Coulomb interactions, respectively. The last term in the total
Hamiltonian (\ref{Ht}) is the light-matter interaction part that is
described in the length-gauge:%
\begin{equation}
\widehat{H}_{\mathrm{int}}=e\sum_{i\sigma }\mathbf{r}_{i}\cdot \mathbf{E}%
\left( t\right) c_{i\sigma }^{\dagger }c_{i\sigma },  \label{Hint}
\end{equation}%
with the elementary charge $e$, position vector $\mathbf{r}_{i}$, and the
electric field strength $\mathbf{E}\left( t\right) $.

In this work, EEI is treated in the HF mean-field approximation
employing the correlation expansion%
\begin{equation*}
\left\langle c_{1}^{\dagger }c_{2}^{\dagger }c_{3}c_{4}\right\rangle
=\left\langle c_{1}^{\dagger }c_{4}\right\rangle \left\langle c_{2}^{\dagger
}c_{3}\right\rangle -\left\langle c_{1}^{\dagger }c_{3}\right\rangle
\left\langle c_{2}^{\dagger }c_{4}\right\rangle .
\end{equation*}%
This factorization technique allows us to obtain a closed set of equations
for the single-particle density matrix $\rho _{ji}^{\left( \sigma \right)
}=\left\langle c_{i\sigma }^{\dagger }c_{j\sigma }\right\rangle $. We will
assume that in the static limit the EEI Hamiltonian vanishes $\widehat{H}_{%
\mathrm{C}}^{HF}\simeq 0$. That is, EEI in the HF limit is
included non-explicitly in empirical TB parameters $\widetilde{\varepsilon }%
_{i}$, $\widetilde{t}_{ij}$ which is chosen to be close to experimental
values. For this propose in this paper we use up to 10th nearest-neighbor
hopping $\widetilde{t}_{ij}$ with values taken from density functional
theory by Wannierization \cite{linhart2018accurate} (see Table 1). Hence,
the Hamiltonian $\widehat{H}_{\mathrm{TB}}+\widehat{H}_{\mathrm{C}}$ is
approximated by,%
\begin{equation*}
\widehat{H}_{0}^{HF}=\sum_{i\sigma }\widetilde{\varepsilon }_{i}c_{i\sigma
}^{\dagger }c_{j\sigma }-\sum_{i,j\sigma }\widetilde{t}_{ij}c_{i\sigma
}^{\dagger }c_{j\sigma }+U\sum_{i}\left( \overline{n}_{i\uparrow }-\overline{%
n}_{0i\uparrow }\right) n_{i\downarrow }
\end{equation*}%
\begin{equation*}
+U\sum_{i\sigma }\left( \overline{n}_{i\downarrow }-\overline{n}%
_{0i\downarrow }\right) n_{i\uparrow }+\sum_{\left\langle i,j\right\rangle
}V_{ij}\left( \overline{n}_{j}-\overline{n}_{0j}\right) n_{i}
\end{equation*}%
\begin{equation}
-\sum_{i,j\sigma }V_{ij}c_{i\sigma }^{\dagger }c_{j\sigma }\left(
\left\langle c_{i\sigma }^{\dagger }c_{j\sigma }\right\rangle -\left\langle
c_{i\sigma }^{\dagger }c_{j\sigma }\right\rangle _{0}\right) ,  \label{HHF}
\end{equation}%
where $\overline{n}_{i\sigma }=\left\langle c_{i\sigma }^{\dagger
}c_{i\sigma }\right\rangle =\rho _{ii}^{\left( \sigma \right) }$. In this
representation the initial density matrix $\rho _{ji}^{\left( \sigma \right)
}\left( 0\right) =\left\langle c_{i\sigma }^{\dagger }c_{j\sigma
}\right\rangle _{0}$ is calculated with respect to renormalized
tight-binding Hamiltonian $\widehat{H}_{0}^{t}=-\sum_{i,j\sigma }\widetilde{t%
}_{ij}c_{i\sigma }^{\dagger }c_{j\sigma }$. From the Heisenberg equation we
obtain evolutionary equations for the single-particle density matrix $\rho
_{ij}^{\left( \sigma \right) }=\left\langle c_{j\sigma }^{\dagger
}c_{i\sigma }\right\rangle $: 
\begin{equation*}
i\hbar \frac{\partial \rho _{ij}^{\left( \sigma \right) }}{\partial t}%
=\sum_{k}\left( \tau _{kj\sigma }\rho _{ik}^{\left( \sigma \right) }-\tau
_{ik\sigma }\rho _{kj}^{\left( \sigma \right) }\right) +\left( V_{i\sigma
}-V_{j\sigma }\right) \rho _{ij}^{\left( \sigma \right) }
\end{equation*}%
\begin{equation}
+e\mathbf{E}\left( t\right) \left( \mathbf{r}_{i}-\mathbf{r}_{j}\right) \rho
_{ij}^{\left( \sigma \right) }-i\hbar \gamma \left( \rho _{ij}^{\left(
\sigma \right) }-\rho _{0ij}^{\left( \sigma \right) }\right) ,  \label{evEqs}
\end{equation}%
where 
\begin{equation}
V_{i\sigma }=\sum_{j\alpha }V_{ij}\left( \rho _{jj}^{\left( \alpha \right)
}-\rho _{0jj}^{\left( \alpha \right) }\right) +U\left( \rho _{ii}^{\left( 
\overline{\sigma }\right) }-\rho _{0ii}^{\left( \overline{\sigma }\right)
}\right) ,  \label{Vij}
\end{equation}%
\ and 
\begin{equation}
\tau _{ij\sigma }=-\widetilde{\varepsilon }_{i}\delta _{ij}+\widetilde{t}%
_{ij}+V_{ij}\left( \rho _{ji}^{\left( \sigma \right) }-\rho _{0ji}^{\left(
\sigma \right) }\right) .  \label{tauij}
\end{equation}%
Electron-electron, electron-phonon scattering processes have been introduced
in Eq. (\ref{evEqs}) phenomenologically via damping term, assuming that the
system relaxes at a rate $\gamma $\ to the equilibrium $\rho _{0ij}^{\left(
\sigma \right) }$\ distribution.

\begin{table*}[tbp]
\caption{The first row represents $n$th-nearest-neighbor order. The second
row is the set of tight-binding parameters, where $t_{ii}=\widetilde{\protect%
\epsilon }_{0}$. The third row is the Coulomb interaction matrix elements.
The first three elements, where $U=V_{ii},$ are obtained from numerical
calculations by using Slater $\protect\pi _{z}$ orbitals \protect\cite%
{potasz2010spin,gucclu2014graphene}. The longer range Coulomb interaction is
taken to be $\protect{\epsilon}_{d} V_{ij}=14.4/d_{ij}$ $eV$, where $d_{ij}$ is
the distance in angstrom between the distant neighbors. Here ${\protect%
\epsilon}_{d} $ is an effective dielectric constant which accounts for the
substrate-induced screening in the 2D nanostructure.}%
\begin{tabular}{cccccccccccc}
\hline\hline
nearest-neighbor & $0$ & $1$ & $2$ & $3$ & $4$ & $5$ & $6$ & $7$ & $8$ & $9$
& $10$ \\ \hline
$t_{ij}\ [\mathrm{eV}]$ & $0.297$ & $2.912$ & $-0.223$ & $0.289$ & $-0.025$
& $-0.055$ & $0.022$ & $0.013$ & $0.022$ & $-0.007$ & $-0.004$ \\ \hline
${\epsilon}_{d} V_{ij}\ [\mathrm{eV}]$ & $16.5$ & $8.6$ & $5.3$ & $14.4/d_{i3}$ & $%
14.4/d_{i4}$ & $14.4/d_{i5}$ & $14.4/d_{i6}$ & $14.4/d_{i7}$ & $14.4/d_{i8}$
& $14.4/d_{i9}$ & $14.4/d_{i10}$ \\ 
&  &  &  &  &  &  &  &  &  &  & 
\end{tabular}%
\end{table*}

In the present paper, as a first approximation, mono-vacancy is simulated by
setting the hopping parameters to the empty site to zero and the on-site
energy at the empty site equals to a large value outside the energy range of
the density of states \cite{pereira2008additional}. There is also scenario
when the structure undergoes a bond reconstruction in the vicinity of the
vacancy \cite{ding2005theoretical}. In either case, a local distortion of
the lattice takes place resulting states that are strongly localized around
defects \cite{pereira2008modeling,lee2005diffusion}. In the tight-binding
Hamiltonian (\ref{HHF}) the diagonal disorder is described by the Anderson
model introducing randomly distributed site energies $\epsilon _{ri}$: $%
\widetilde{\epsilon }_{i}=\widetilde{\epsilon }_{0i}+\epsilon _{ri}$. We
assume for the random variable $\epsilon _{ri}$ to have probability
distributions $P\left( \epsilon _{ri},V_{\mathrm{on}}\right) $, where 
\begin{equation}
P\left( \epsilon _{ri},V_{\mathrm{on}}\right) =\left\{ 
\begin{array}{c}
\frac{1}{2V_{\mathrm{on}}},\ -V_{\mathrm{on}}\leq \epsilon _{ri}\leq V_{%
\mathrm{on}} \\ 
0,\ \mathrm{otherwise}%
\end{array}%
\right. .  \label{dist}
\end{equation}%
Here the quantity $V_{\mathrm{on}}$ is the distribution width describing the
strength of the disorder. The disorder strength for all calculations is
taken to be $V_{\mathrm{on}}$=0.3 eV.

\section{Linear optical response}

The linear optical response of the considered system is completely described
by the initial equilibrium $\rho _{0ij}^{\left( \sigma \right) }$\
distribution function. For this propose we need the eigenfunctions ($\psi
_{\sigma \mu }\left( i\right) $) and eigenenergies ($\varepsilon _{\sigma
\mu }$) of the TB Hamiltonian. We numerically diagonalize the TB Hamiltonian
with the parameters from Table 1, and construct the initial density matrix $%
\rho _{0ij}^{\left( \sigma \right) }$ via the filling of electron states in
the valence band according to the zero temperature Fermi--Dirac-distribution 
$\rho _{0ij}^{\left( \sigma \right) }=\sum_{\mu =N/2}^{N-1}\psi _{\sigma \mu
}^{\ast }\left( j\right) \psi _{\sigma \mu }\left( i\right) $. In Fig. 1 the
electron probability density $\left\vert \psi _{\sigma \mu }\left( i\right)
\right\vert ^{2}$ on the 2D color mapped nanostructure corresponding to the
highest energy level in the valence band and eigenenergies near the Fermi
level are shown. In both cases we see the emergence of states near the Fermi
level. As is also seen from this figure, the presence of on-site disorder or
a mono-vacancy breaks the inversion symmetry. As expected, in the case of a
vacancy we have a state that is strongly localized around the defect. To
provide a quantitative characterization of localization in the $\mu $-th
eigenstates, we also calculate the inverse participation number%
\begin{equation}
P_{\mu }=\left( \sum_{i=0}^{N-1}\left\vert \psi _{\mu }\left( i\right)
\right\vert ^{4}\right) ^{-1},  \label{IPN}
\end{equation}%
which provides a measure of the fraction of sites over which the wavepacket
is spread \cite{kramer1993localization}. The normalized inverse
participation numbers $P_{\mu }/N$ for states near the Fermi level are
illustrated in Fig. 2. From this figure it becomes evident that these states
are localized, which as we will see, strongly alter the optical response of
the considered nanostructures. For the higher energy states $P_{\mu }/N\sim
0.5$, which means that those states are more resistant to defects. 
\begin{figure}[tbp]
\includegraphics[width=0.46\textwidth]{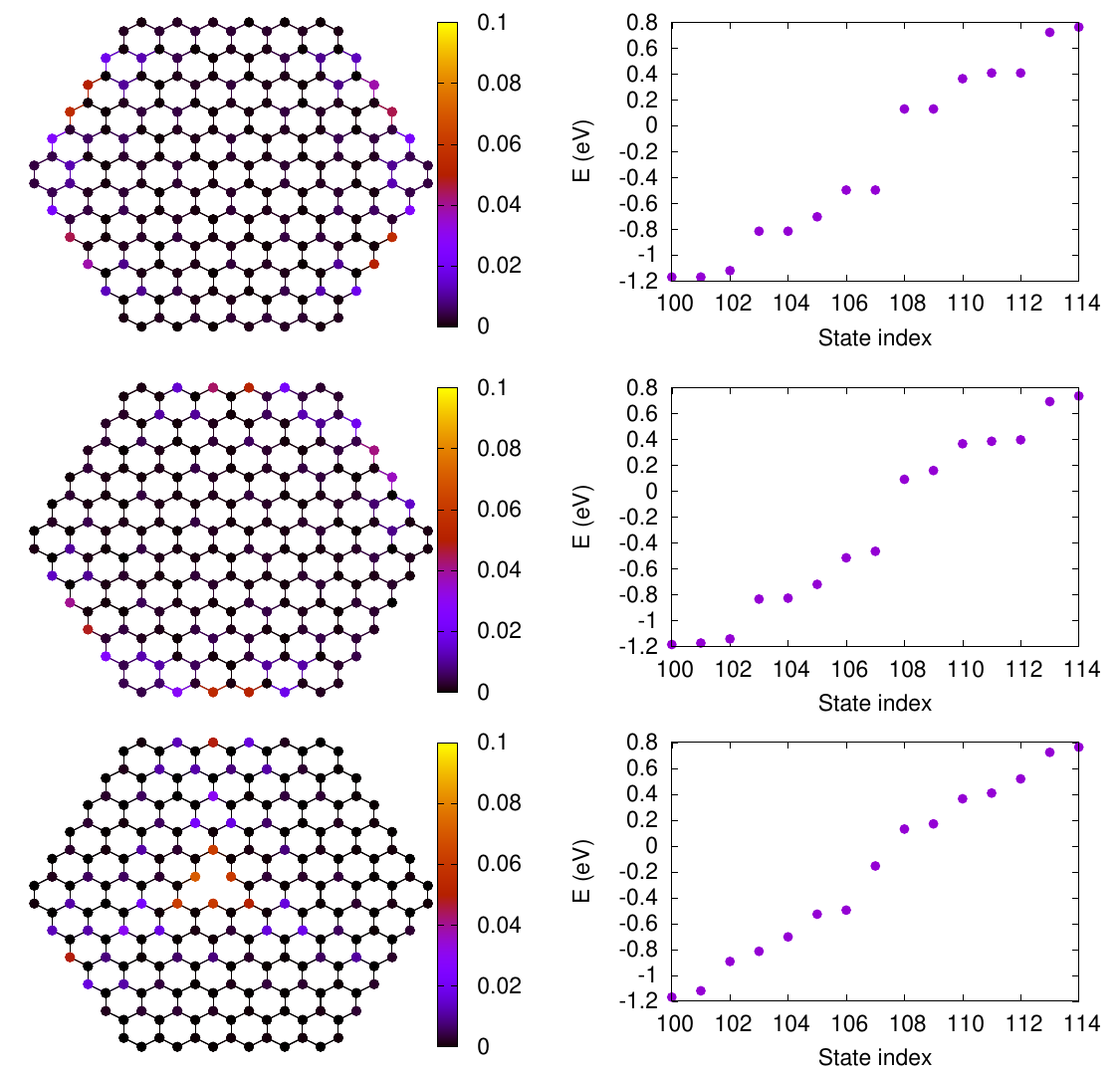}
\caption{The top row represents intrinsic GQD$_{216}$, the middle row
depicts GQD$_{216}$ with disorder, and the bottom row illustrates GQD$_{216}$
with a monovacancy. Within each row, the following visualizations are
presented from left to right: electron probability density corresponding to
the highest energy level in the valence band and eigenenergies near the
Fermi level.}
\end{figure}
\begin{figure}[tbp]
\includegraphics[width=0.35\textwidth]{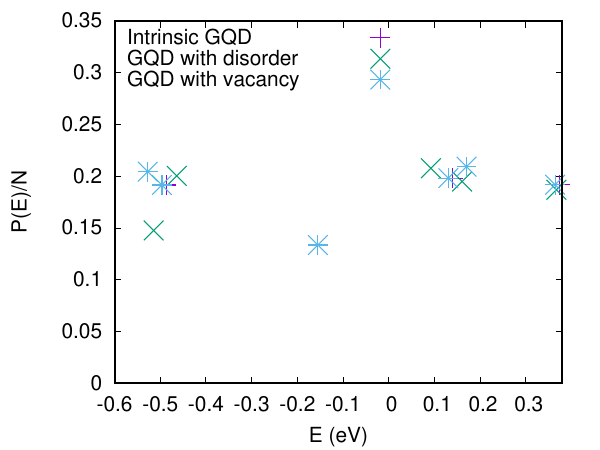}
\caption{The normalized inverse participation number for states near the
Fermi level.}
\end{figure}

To study the linear response we first need to calculate the susceptibility
tensor $\chi _{ij}$ which is more transparent to give in the energetic
representation. For this propose we perform a basis transformation using the
following formula:%
\begin{equation}
\rho _{ij}^{(\sigma )}=\sum_{\mu ^{\prime }}\sum_{\mu }\psi _{\sigma \mu
^{\prime }}^{\ast }\left( j\right) \varrho _{\sigma \mu \mu ^{\prime }}\psi
_{\sigma \mu }\left( i\right) ,  \label{2}
\end{equation}%
where $\varrho _{\sigma \mu \mu ^{\prime }}$\textrm{\ }is the density matrix
in the energetic representation. Taking into account the completeness of
basis functions, from Eq. (\ref{evEqs}) we get the following equation 
\begin{equation*}
i\hbar \frac{\partial \varrho _{\sigma mn}}{\partial t}=\left( \varepsilon
_{\sigma m}-\varepsilon _{\sigma n}\right) \varrho _{\sigma mn}
\end{equation*}%
\begin{equation}
+\mathbf{E}\left( t\right) \sum_{\mu }\left( \varrho _{\sigma \mu n}\mathbf{d%
}_{\sigma m\mu }-\varrho _{\sigma m\mu }\mathbf{d}_{\sigma \mu n}\right)
-i\hbar \gamma \left( \varrho _{\sigma mn}-\varrho _{\sigma mn}^{(0)}\right)
,  \label{main}
\end{equation}%
where $\mathbf{d}_{\sigma \mu ^{\prime }\mu }=e\sum_{i}\psi _{\sigma \mu
^{\prime }}^{\ast }\left( i\right) \mathbf{r}_{i}\psi _{\sigma \mu }\left(
i\right) $ is the transition dipole moment. We will solve Eq. (\ref{main})
in the scope of perturbation theory by expanding the density matrix in
orders of the incident electromagnetic field%
\begin{equation}
\varrho _{\sigma mn}=\varrho _{\sigma mn}^{(0)}+\varrho _{\sigma mn}^{(1)}.
\label{pert}
\end{equation}%
Assuming $E_{j}\left( t\right) =\sum_{\omega }E_{j}\left( \omega \right)
e^{-i\omega t}$, it is straightforward to obtain 
\begin{equation*}
\varrho _{\sigma mn}^{(1)}=\sum_{j\omega }\frac{E_{j}\left( \omega \right)
e^{-i\omega t}d_{\sigma mn}^{j}\left( \varrho _{\sigma nn}^{(0)}-\varrho
_{\sigma mm}^{(0)}\right) }{\varepsilon _{\sigma m}-\varepsilon _{\sigma
n}-\hbar \omega -i\hbar \gamma }.
\end{equation*}%
The polarization vector $P_{i}\left( t\right) =\sum_{\sigma mn}\varrho
_{\sigma mn}^{(1)}\left( t\right) d_{\sigma nm}^{i}$ now can be expressed
as: 
\begin{equation}
P_{i}\left( t\right) =\sum_{\sigma mn}\sum_{j\omega }\frac{E_{j}\left(
\omega \right) e^{-i\omega t}d_{\sigma mn}^{j}\left( \varrho _{\sigma
nn}^{(0)}-\varrho _{\sigma mm}^{(0)}\right) }{\varepsilon _{\sigma
m}-\varepsilon _{\sigma n}-\hbar \omega -i\hbar \gamma }.  \label{Pt}
\end{equation}%
Taking into account the definition $P_{i}=\epsilon _{0}\sum_{j}\chi
_{ij}E_{j}$, where $\epsilon _{0}$ is the electric permittivity of free
space, from Eq. (\ref{Pt}) we get: 
\begin{equation*}
\chi _{ij}\left( \omega \right) =\frac{1}{\epsilon _{0}}\sum_{\sigma
mn}\varrho _{\sigma mm}^{(0)}\left[ \frac{d_{\sigma nm}^{j}d_{\sigma mn}^{i}%
}{\varepsilon _{\sigma n}-\varepsilon _{\sigma m}-\hbar \omega -i\hbar
\gamma }\right.
\end{equation*}%
\begin{equation}
\left. +\frac{d_{\sigma mn}^{j}d_{\sigma nm}^{i}}{\varepsilon _{\sigma
n}-\varepsilon _{\sigma m}+\hbar \omega +i\hbar \gamma }\right] .
\label{kij}
\end{equation}%
With the help of susceptibility tensors in SI units we also calculate the
conductivity tensors in CGS units by the formula%
\begin{equation}
\sigma _{ij}\left( \omega \right) =-i\epsilon _{0}\omega \chi _{ij}\left(
\omega \right) .  \label{CI1}
\end{equation}%
\begin{figure}[tbp]
\includegraphics[width=0.46\textwidth]{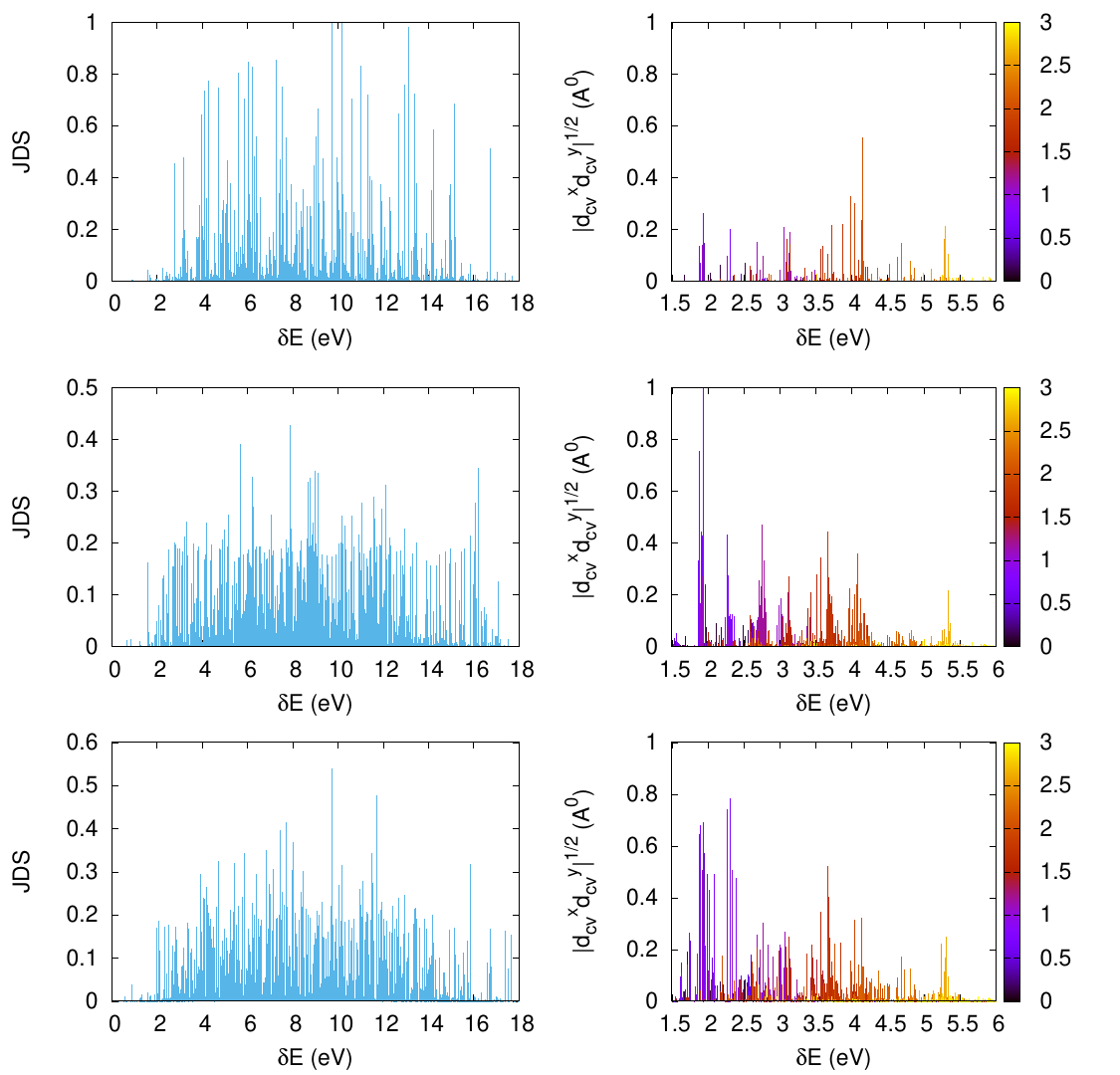}
\caption{The top panel represents intrinsic GQD$_{216}$, the middle panel
depicts GQD$_{216}$ with disorder, and the bottom panel illustrates GQD$%
_{216}$ with a monovacancy. The left column is the joint density of states,
and the right column is the absolute values of the product of the $x$ and $y$
components of interband transition dipole matrix elements. The color boxes
indicate the energy ranges (in eV) of the conduction bands.}
\end{figure}
\qquad

The tensor $\sigma _{ij}\left( \omega \right) $ can be dissected into
symmetric and antisymmetric components. For 2D systems, there exists a
single independent component that characterizes the antisymmetric
conductivity: $\sigma _{xy}\left( \omega \right) =-\sigma _{yx}\left( \omega
\right) $. This component is commonly known as the dissipationless or Hall
conductivity, denoted as $\sigma _{H}\left( \omega \right) $. Under spatial
symmetry transformations, the Hall component transforms as a pseudo-scalar 
\cite{nandy2019symmetry}, given by $\sigma _{H}\left( \omega \right) =\det
\left( O\right) \sigma _{H}\left( \omega \right) $. Consequently, the Hall
conductivity becomes zero in systems with mirror symmetry, where $\det
\left( O\right) =-1$. Furthermore, it also vanishes in time-reversal
invariant systems. To observe a non-zero Hall component in the considered
GQD, it is imperative to break both time-reversal and inversion symmetries.
The time-reversal symmetry can be disrupted, for example, through the
application of a magnetic field. Remarkably, in the context of GQDs
characterized by sharp boundaries, the time-reversal symmetry is inherently
broken even in the absence of an external magnetic field \cite%
{berry1987neutrino}. This intriguing property hints at the potential for
Hall-type anomalous responses in such imperfect GQDs. For the purposes of
our analysis, we neglect spin effects and assume a zero-temperature
Fermi-Dirac distribution for initial density matrix $\varrho _{\sigma
mm}^{(0)}$. Under these conditions, the expressions for the conduction
tensor can be derived from Eqs. (\ref{kij}) and (\ref{CI1}) as follows:%
\begin{equation*}
\sigma _{ij}\left( \omega \right) =-2i\omega \sum_{m\in \nu }\sum_{n\in
c}d_{nm}^{j}d_{mn}^{i}
\end{equation*}%
\begin{equation}
\times \left[ \frac{1}{\varepsilon _{n}-\varepsilon _{m}-\hbar \omega
-i\hbar \gamma }+\frac{1}{\varepsilon _{n}-\varepsilon _{m}+\hbar \omega
+i\hbar \gamma }\right] .  \label{sij}
\end{equation}%
As evident from Eq. (\ref{sij}), the conduction tensor is primarily defined
by the joint density of states (JDS):%
\begin{equation*}
JDS=\sum_{m\in \nu }\sum_{n\in c}\delta \left( \varepsilon _{n}-\varepsilon
_{m}-\hbar \omega \right)
\end{equation*}%
and the values of the product $d_{nm}^{j}d_{mn}^{i}$ of the interband
transition dipole matrix elements. In Fig. 3, we illustrate the JDS and the
absolute values of the product of the $x$ and $y$ components of the
interband transition dipole matrix elements. As is seen from this figure,
the influence of on-site disorder and a mono-vacancy on GQD$_{216}$ is
nearly identical. While the JDS is somewhat suppressed, both factors lift
the degeneracy of states and break the inversion symmetry, thereby opening
up new channels for interband transitions. The product $d_{cv}^{x}d_{cv}^{y}$
has several peaks near the particular interband transitions, which means
that near these peaks one can expect strong Hall-type anomalous response. To
assess the consequences on the linear response using $\sigma _{ij}\left(
\omega \right) $, we will compute the linear absorption coefficient and the
Faraday rotation angle. For both quantities, we will use formulas derived
for graphene at normal incidence of a laser beam. The linear absorption
coefficient is defined through the diagonal component of conductivity \cite%
{stauber2008optical}:%
\begin{equation}
\alpha _{\mathrm{abs}}=\frac{4\pi }{c}\frac{\mathrm{Re}\sigma _{xx}\left(
\omega \right) }{\left\vert \frac{\sqrt{\epsilon }+1}{2}+2\pi \sigma
_{xx}\left( \omega \right) /c\right\vert ^{2}},  \label{abs}
\end{equation}%
while the Faraday-rotation angle $\Theta _{F}$ is related to the optical
Hall conductivity \cite{morimoto2010optical} through the formula:%
\begin{equation}
\Theta _{F}=\frac{1}{2}arg\left[ \frac{1+\sqrt{\epsilon }+\frac{4\pi }{c}%
\left( \sigma _{xx}\left( \omega \right) +i\sigma _{xy}\left( \omega \right)
\right) }{1+\sqrt{\epsilon }+\frac{4\pi }{c}\left( \sigma _{xx}\left( \omega
\right) -i\sigma _{xy}\left( \omega \right) \right) }\right] .  \label{Farad}
\end{equation}%
Here, $c$ represents the velocity of light, and $\epsilon $ denotes the
dielectric constant of the substrate. Strictly speaking, the absorption
coefficient and Faraday rotation angle are meaningful for a nanostructure
layer with dimensions much larger than the incident light wavelength. In
other words, we should have many copies of the GQDs uniformly distributed on
a 2D surface. We investigate these quantities to emphasize the consequences
of disorder or vacancies on the optical response of GQDs. In Fig. 4, we
present the linear optical response of GQD$_{216}$ in terms of the
absorption coefficient and Faraday rotation angle. Notably, the absorption
coefficient is minimally affected by disorder or vacancies. In comparison to
graphene, where $\alpha _{\mathrm{abs}}\approx {\pi }/137$, we observe a
significantly higher absorption ($\alpha _{\mathrm{abs}}\sim 0.1$). In terms
of Hall-type anomalous response, we observe a substantial Faraday-rotation
angle for mono-vacancy and a less pronounced effect for disorder. As
expected, intrinsic GQD displays a Faraday-rotation angle of zero. It is
worth noting that the maximal angle induced by a mono-vacancy, $\Theta
_{F}\sim 3^{%
{{}^\circ}%
}$, is comparable to the Faraday-rotation angle of graphene in the strong
magnetic field with strengths $\sim 3\cdot 10^{4}$ Gs \cite%
{ferreira2011faraday}. 
\begin{figure}[tbp]
\includegraphics[width=0.5\textwidth]{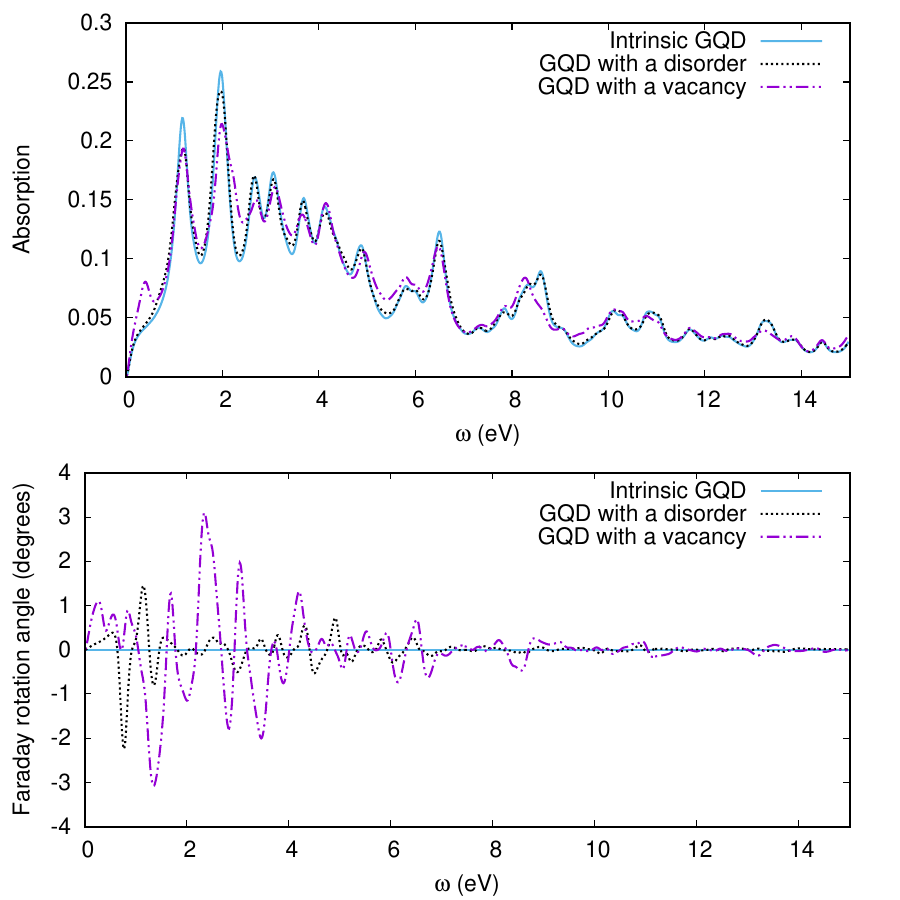}
\caption{Linear optical response of GQD via absorption coefficient (a) and
Faraday rotation angle (b). The relaxation rate is set to $\hbar \protect%
\gamma =0.1\ \mathrm{eV}$. The dielectric constant of the substrate is taken
to be $\protect{\epsilon}_{d} =6$. }
\end{figure}

\section{Nonlinear optical response}

After considering the linear response, we begin by examining the nonlinear
optical response of GQD$_{216}$ in the strong infrared laser field described
by the electric field strength $\mathbf{E}\left( t\right) =f\left( t\right)
E_{0}\hat{\mathbf{e}}\cos \omega t$, with the frequency $\omega $,
polarization $\hat{\mathbf{e}}$ unit vector, and amplitude $E_{0}$. The wave
envelope is described by the Gaussian function $f\left( t\right) =\exp \left[
-2\ln 2\left( t-t_{m}\right) ^{2}/\mathcal{T}^{2}\right] $, where $\mathcal{T%
}$ characterizes the pulse duration full width at half maximum, $t_{m}$
defines the position of the pulse maximum. Note that for the Gaussian
envelope the number of oscillations $N_{s}$ of the field is approximated as $%
\mathcal{T}/T\simeq 0.307N_{s}$, where $T=2\pi /\omega $ is the wave period (%
$4.135\mathrm{fs}$ for 1$\mathrm{eV}$).

To compute the harmonic signal, we use the Fourier transform 
\begin{equation}
\mathbf{a}\left( \Omega \right) =\int_{-\infty }^{\infty }\mathbf{a}\left(
t\right) e^{i\Omega t}W\left( t\right) dt,  \label{daccw}
\end{equation}%
where 
\begin{equation}
\mathbf{a}\left( t\right) =e\sum_{i\sigma }\mathbf{r}_{i}\frac{d^{2}}{dt^{2}}%
\rho _{ii}^{\left( \sigma \right) }\left( t\right)  \label{dacc}
\end{equation}%
is the dipole acceleration and $W\left( t\right) $ is a window function that
suppresses small fluctuations and reduces the overall background noise of
the harmonic signal \cite{zhang2018generating}. We choose the pulse envelope 
$f\left( t\right) $ as the window function. For all further calculations we
assume a polarization unit vector $\hat{\mathbf{e}}=\left\{ 1,0\right\} $,
and the pulse duration $\mathcal{T}$ is set to $\mathcal{T}/T\simeq 3$,
corresponding to approximately $10$ oscillations ($N_{s}\simeq 10$). To
ensure a smooth turn-on of the interaction, we position the pulse center at $%
t_{m}=10T$. For convenience, we normalize the dipole acceleration by the
factor $a_{0}=e\overline{\omega }^{2}\overline{d},$ where $\overline{\omega }%
=1\ \mathrm{eV}/\hbar $ and $\overline{d}=1\ \mathrm{\mathring{A}}$. The
power radiated at a given frequency is proportional to $S\left( \Omega
\right) =\left\vert \mathbf{a}\left( \Omega \right) \right\vert
^{2}/a_{0}^{2}$. We perform the time integration of Eq. (\ref{evEqs}) using
the eighth-order Runge-Kutta algorithm. For the Coulomb interaction matrix
elements we take values from Table 1 and the dielectric constant of the
substrate is taken to be ${\epsilon}_{d} =6$. 
\begin{figure}[tbp]
\includegraphics[width=0.47\textwidth]{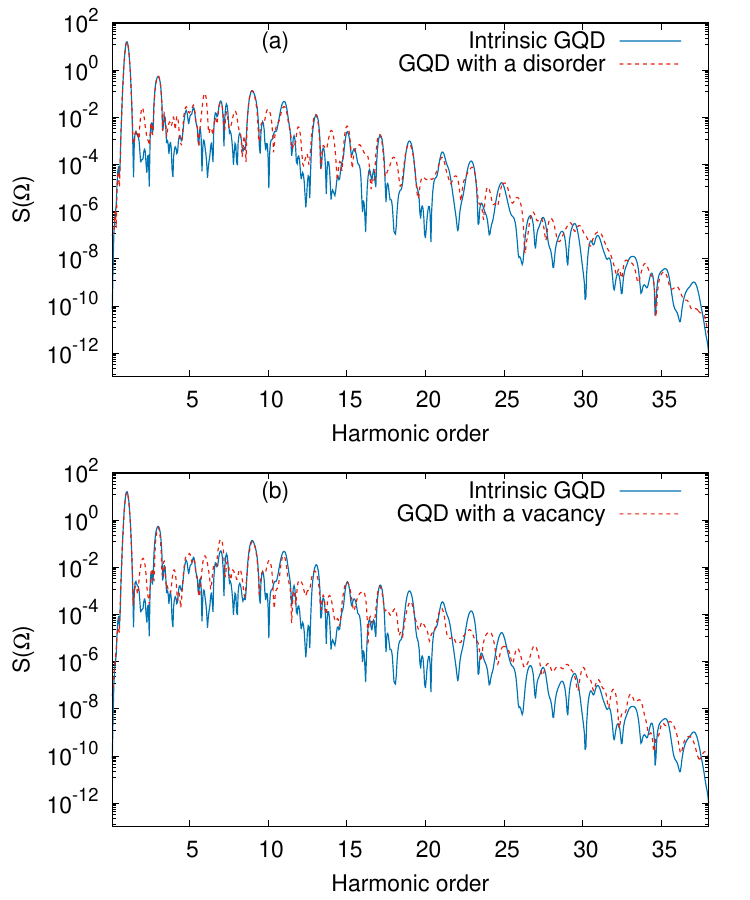}
\caption{The total HHG spectra in logarithmic scale (in arbitrary units) for
GQD$_{216}$ with a disorder (a) and for GQD$_{216}$ with monovacancy (b)
along with the HHG spectra of intrinsic GQD. The wave amplitude is taken to
be $E_{0}=0.3\ \mathrm{V/\mathring{A}}$. The relaxation rate is set to $%
\hbar \protect\gamma =0.1\ \mathrm{eV}$. The excitation is performed
assuming a laser with a wavelength of $2.48\ \mathrm{\protect\mu m}$, an
excitation frequency of $\protect\omega =0.5\ \mathrm{eV}/\hbar $.}
\end{figure}

To begin with, we examine the effect of the mono-vacancy and disorder on the
HHG spectra. The HHG spectra are compared for three different scenarios in
Fig. 5: when we have the intrinsic GQD$_{216}$, when the GQD$_{216}$ has a
mono-vacancy, and when it is subject to on-site disorder. The inclusion of a
mono-vacancy or disorder leads to two noteworthy characteristics in the HHG
spectra: (a) the most prominent feature is the emergence of even harmonics
comparable to odd harmonics and (b) substantial increase in the HHG signal
in the vicinity of the cutoff regime. The first phenomenon is attributed to
the special inversion symmetry breaking in the presence of disorder and
vacancy. The second phenomenon is connected with the fact that the disorder
and vacancy lift the degeneracy of states opening up new channels for
interband transitions.

To reveal the inherent effects of broken time-reversal symmetry on the
nonlinear response, we conducted an investigation into polarization-resolved
HHG spectra. In Fig. 6, we present the polarization-resolved HHG spectra for
GQD$_{216}$ with disorder and for GQD$_{216}$ featuring a mono-vacancy. This
figure highlights a significant finding, especially in the case of a
mono-vacancy, where even-order harmonics manifest as Hall-type anomalous
harmonics polarized perpendicular to the applied laser electric field
direction. 
\begin{figure}[tbp]
\includegraphics[width=0.47\textwidth]{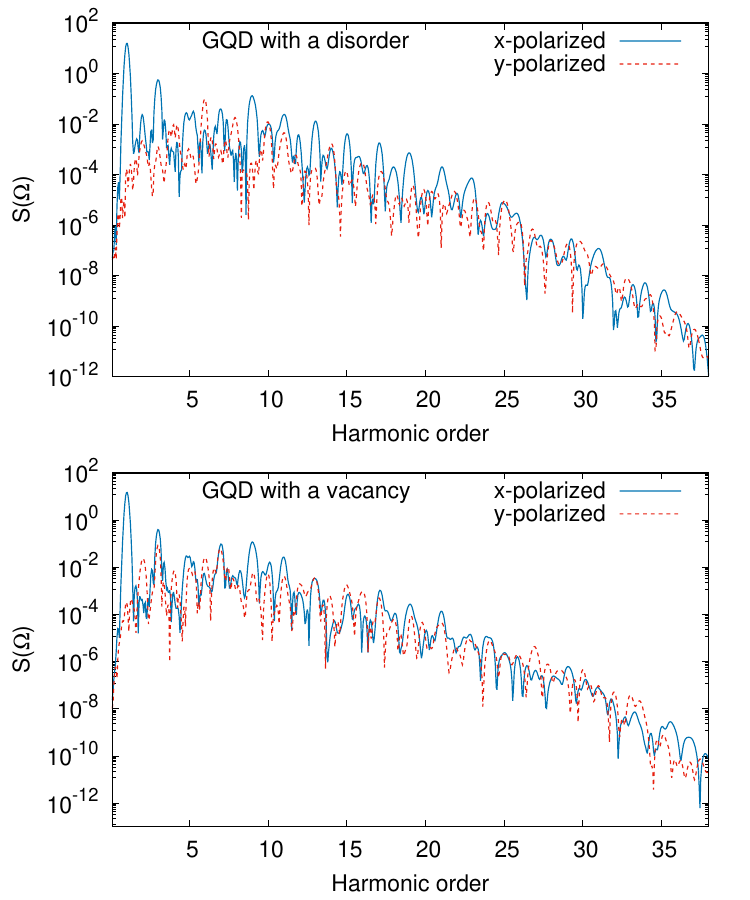}
\caption{The polarization resolved HHG spectra in logarithmic scale for GQD$%
_{216}$ with a disorder (a) and for GQD$_{216}$ with a monovacancy (b). The
laser parameters correspond to Fig. 5.}
\end{figure}

Of specific interest is the plateau region within the harmonics spectra. In
Fig. 7, we present the plateau portion of the anomalous HHG spectrum of GQD$%
_{216}$ with a mono-vacancy for various wave field amplitudes. In this
representation, none of the harmonics conform to the perturbation scaling $%
S^{1/2}\left( n\omega \right) \sim E_{0}^{n}$. This observation underscores the
strictly multiphoton and nonlinear nature of the HHG process. 
\begin{figure}[tbp]
\includegraphics[width=0.47\textwidth]{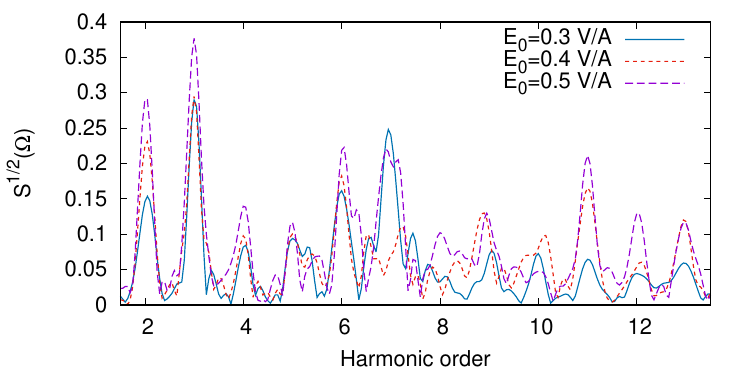}
\caption{The plateau part of the anomalous HHG spectrum of GQD$_{216}$ with
a monovacancy is presented in the linear scale for various wave field
amplitudes. The excitation frequency is taken to be $\protect\omega =0.5\ 
\mathrm{eV}/\hbar $. }
\end{figure}

Now, let's consider the effect of the pump wave frequency on the Hall-type
anomalous HHG process. This analysis is presented in Fig. 8 where we
demonstrate the polarization-resolved HHG spectra for higher-frequency laser
fields. Notably, we observe that the rate of anomalous harmonics is
suppressed for higher-frequency pump waves. This phenomenon can be
attributed to the fact that with higher-frequency pump waves, excitation and
recombination channels predominantly involve highly excited states that
still retain inversion symmetry. 
\begin{figure}[tbp]
\includegraphics[width=0.47\textwidth]{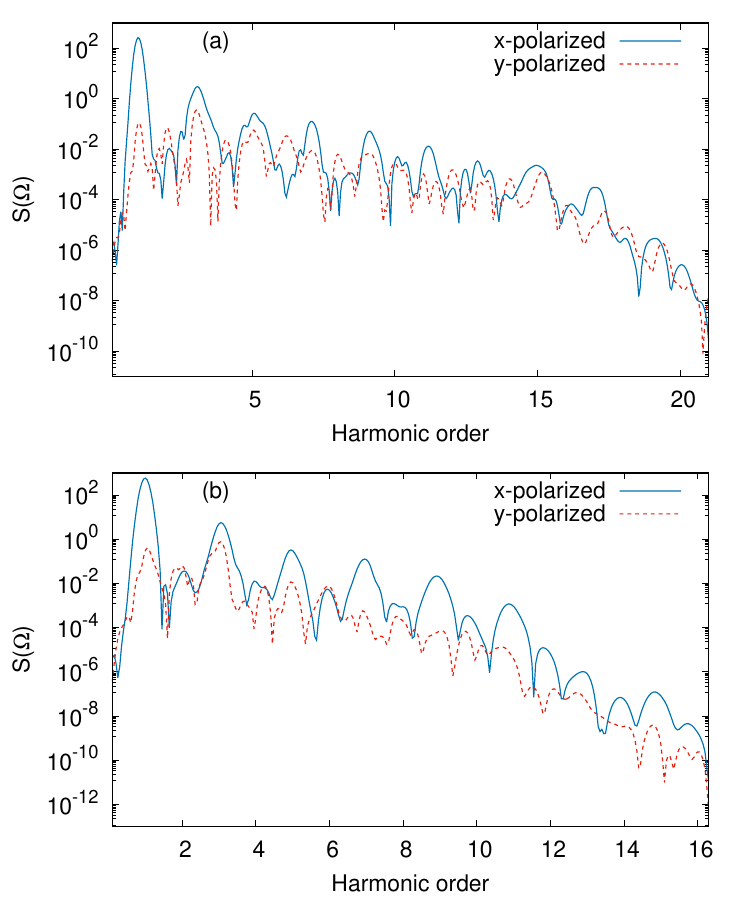}
\caption{The polarization resolved HHG spectra in logarithmic scale for GQD$%
_{216}$ with a monovacancy for $\protect\omega =1.0\ \mathrm{eV}/\hbar $ (a)
and $\protect\omega =1.5\ \mathrm{eV}/\hbar $ (b). The wave amplitude is
taken to be $E_{0}=0.3\ \mathrm{V/\mathring{A}}$. The relaxation rate is set
to $\hbar \protect\gamma =0.1\ \mathrm{eV}$.}
\end{figure}

\section{Conclusion}

We have studied the character and specifics of linear and nonlinear optical
response of hexagonal GQDs in the presence of on-site disorder or vacancies.
Our primary focus was on medium-sized GQDs composed of $216$ carbon atoms
which in their defectless state possess inversion symmetry. To model a
disorder, we employed the Anderson model, while for mono-vacancies we
utilized a simplified model by setting the hopping parameters to the empty
site to zero and assigning a large value to the on-site energy at the empty
site. In our TB model, we considered up to the $10$th nearest-neighbor
hopping elements. Electron-electron interactions were treated within the
HF approximation, incorporating the long-range Coulomb
interactions. Through the solution of the evolutionary equations for the
single-particle density matrix, we revealed anomalous optical responses in
defective GQDs across both linear and nonlinear interaction regimes. In
linear response, in the absence of an external magnetic field, we observed a
significant Hall conductivity, resulting in a substantial Faraday-rotation
angle. This phenomenon is attributed to the intrinsic time-reversal
symmetry-breaking in graphene quantum dots, coupled with the simultaneous
breaking of spatial inversion symmetry in the presence of disorder or
mono-vacancies. The combined disruption of time-reversal and inversion
symmetries leads to the emergence of intense Hall-type anomalous HHG.
Notably, these anomalous high harmonics are more pronounced in low-frequency
laser fields. In such scenarios, excitation and recombination channels
predominantly involve states near the Fermi level which are more susceptible
to inversion symmetry breaking. Our findings underscore that even minor
levels of disorder or mono-vacancy leave unique imprints in the Hall-type
anomalous high harmonics, thus offering a potential avenue for optically
characterizing defects in 2D nanostructures.

\begin{acknowledgments}
The work was supported by the Science Committee of Republic of
Armenia, project No. 21AG-1C014.
\end{acknowledgments}

\bibliographystyle{apsrev4-2}
\bibliography{bibliography}

\end{document}